\documentclass[twocolumn,pra,showpacs,preprintnumbers,superscriptaddress,amsmath,amssymb,10pt,aps,floatfix]{revtex4}

\usepackage{mathptmx}
\usepackage{subfigure}
\usepackage{psfrag,graphicx}
\usepackage{dcolumn}
\usepackage{bm}
\usepackage{color}
\usepackage{epstopdf}
\usepackage{color}
\usepackage[english]{babel}
\usepackage{latexsym}
\usepackage{psfrag,graphicx}
\usepackage{subfigure}
\usepackage{amsmath}
\usepackage{amssymb}
\usepackage{amsfonts}
\usepackage{bm}
\usepackage{natbib}
\usepackage{epstopdf}
\usepackage{appendix}

\definecolor{mygrey}{gray}{0.35}
\definecolor{myblue}{rgb}{0.2,0.2,0.8}
\definecolor{myzard}{cmyk}{0,0,0.05,0}
\definecolor{mywhite}{rgb}{1,1,1}
\definecolor{mywhite}{rgb}{1,1,1}
\definecolor{myred}{rgb}{1,0.,0.3}

\usepackage[colorlinks=true,citecolor=myblue,linkcolor=myred]{hyperref}

\def\ba{\begin{align}}
\def\enda{\end{align}}
\def\bi{\begin{itemize}}
\def\ei{\end{itemize}}

\def\be{\begin{equation}}
\def\ee{\end{equation}}
\def\bea{\begin{eqnarray}}
\def\eea{\end{eqnarray}}
\def\bse{\begin{subequations}}
\def\ese{\end{subequations}}

\begin{document}

\title{Quantum chaos and thermalization in the two-mode Dicke model}

\author{Aleksandrina V. Kirkova}
\affiliation{Department of Physics, St. Kliment Ohridski University of Sofia, James Bourchier 5 blvd, 1164 Sofia, Bulgaria}

\author{Peter A. Ivanov}
\affiliation{Department of Physics, St. Kliment Ohridski University of Sofia, James Bourchier 5 blvd, 1164 Sofia, Bulgaria}

\begin{abstract}
We discuss the onset of quantum chaos and thermalization in the two-mode Dicke model, which describes the dipolar interaction between an ensemble of spins and two bosonic modes. The two-mode Dicke model exhibits normal to superradiant quantum phase transition with spontaneous breaking either of a discrete or continuous symmetry. We study the behaviour of the fidelity out-of-time-order correlator derived from the Loschmidt echo signal in the quantum phases of the model and show that its exponential growth cannot be related to a classical unstable point in the general case. Moreover, we find that the collective spin observable in the two-mode Dicke model quickly saturates to its long-time average value, and shows very good agreement between its diagonal ensemble average and microcanonical average even for a small number of spins. We show that the temporal fluctuations of the expectation value of the collective spin observable around its average are small and decrease with the effective system size, which leads to thermalization of the spin system.
\end{abstract}

\maketitle

\begin{section}{Introduction}

One characteristic of chaos in quantum many-body systems is the rapid dispersal of quantum information stored in local degrees of freedom among global ones and it becoming effectively lost to all local probes. The apparent loss of information in closed quantum systems is a key mechanism for understanding thermalization of local observables under unitary time evolution. Indeed, a closed quantum system initialized in a pure state remains pure under the action of unitary time evolution, whereas thermalization and emergence of statistical mechanics occur on a local scale \cite{Rigol2008,Eisert2014,Alessio,Gogolin2016}. This is because a large subsystem of the total quantum many-body system can be entangled with the rest of the system and thus can be considered as an effective heat bath, which drives the smaller subsystem towards equilibration and thermalization \cite{Neill2016,Smith2016,Linden2009}. Such a deep connection between quantum information scrambling, entanglement and thermalization has been experimentally explored in various quantum optical systems \cite{Swan2019,Kaufman2016,Clos2016}.

One measure of quantum information scrambling is the double commutator out-of-time-order correlation function (OTOC) \cite{Maldacena2016,Swingle2018,Mata2018}. Its exponentially fast growth in time is characterized by a positive Lyapunov exponent and it is thus associated with the onset of chaos in quantum systems. The OTOC has been regarded as an important observable in the context of the AdS/CFT correspondence and can be used to quantify properties of fast scrambling in black hole dynamics \cite{Shenker2014,Sekino,Markovic2022}. Recently, it was shown that the OTOC can be also used as an indicator of equilibrium as well as non-equilibrium quantum phase transitions \cite{Heyl2018,Shen2017,Swan2020,Garttner2018,Dag2019}. For example the exponential growth of OTOC is related to the onset of a quantum phase transition between the normal and superradiant phases of the Dicke and quantum Rabi models \cite{Swan2019,carlosprl2019,Altland2012,Sun2020,Kirkova2022}. Moreover, measurements of the OTOC have already been performed in several quantum optical systems, including nuclear magnetic resonance (NMR) \cite{Li2017}, superconducting circuits \cite{Braumuller2022}, trapped ions \cite{Garttner2017,Landsman2019,Joshi2020,Green2022}, and cold atoms \cite{Pegahan2021}.

In this work we investigate signatures of chaos and transition to equilibration and thermalization in the two-mode Dicke (TMD) model, which consists of an ensemble of $N$ two-state atoms and two bosonic modes which interact via dipolar coupling. The TMD model shows a quantum phase transition between a normal phase and a superradiant phase \cite{Ivanov2013,Fan,Porras2012,Ivanov2015}. For the $\mathbb{Z}_{2}$-symmetric TMD model the superradiant phase is characterized by macroscopical excitations of one of the bosonic modes, whereas the other mode is with zero mean-field bosonic excitations. For the U(1)-symmetric case, the quantum phase transition is related to the macroscopic excitation of both bosonic modes. We consider the OTOC derived from the Loschmidt echo signal, which is used to quantify imperfect time reversal in a quantum system due to an instantaneously applied perturbation \cite{Schmitt2019}, to study the onset of chaos in the TMD model. Following the current terminology, we refer to it as a fidelity OTOC (FOTOC). For the $\mathbb{Z}_{2}$-symmetric case we find that the OTOC corresponding to the macroscopically excited mode displays exponential growth in the beginning of the time evolution, which then reaches a level of saturation, where it remains for all subsequential times, while oscillating with a small amplitude. We also observe that the OTOC corresponding to the non-excited mode shows slow non-exponential growth with larger temporal oscillations. In the case of the U(1)-symmetric TMD model, the OTOCs corresponding to both modes grow exponentially. Moreover, following \cite{Cameo} we perform a semiclassical analysis of the TMD model and show that it has unstable points, which give rise to classical Lyapunov exponents. We find that the quantum Lyapunov exponent agrees with the largest classical one, provided the latter is non-zero.
Remarkably, we also find a quantum-chaotic parametric regime, where we observe exponential growth of the OTOC, although the corresponding classical Lyapunov exponent is zero.

Furthermore, we study the evolution of observables after a quantum quench. We demonstrate that even for a small number of spins, the collective spin observable approaches quantum thermalization, which allows for the study of ergodic dynamics in a small quantum system. We show that there is very good agreement between the diagonal ensemble average and microcanonical ensemble average of the collective spin observable. The observable quickly reaches its long-time average value as well. Moreover, we show that the long-time average of the temporal fluctuations is small and decreases with the increase of the spin-boson coupling and effective system size \cite{Nation2019}. Thus, the two bosonic degrees of freedom can be regarded as an effective bath system, which drives the collective spin observable towards quantum thermalization.

Finally, we show that the TMD model can be implemented with trapped ions \cite{Wineland1998,Schneider2012}. The two bosonic modes are represented by the radial center-of-mass collective modes and the collective spins are formed by the internal electronic states of the ions. The laser fields along the two orthogonal spatial directions generate the spin-phonon couplings. For the typical ion trap parameters such as spin-phonon coupling and an effective bosonic frequency, we show that quantum thermalization can be observed in $\mu$s regime.

The paper is organized as follows: In Sec. \ref{TMD} we introduce the TMD model and its symmetries. Depending on the parameter regime, the TMD model can exhibit a quantum phase transition, which is associated with symmetry breaking either of a discrete parity symmetry or a continuous U(1) symmetry. In Sec. \ref{fotoc_sec} we study the onset of quantum chaos in the TMD model by investigating the behaviour of the fidelity out-of-time-order correlator in its quantum phases. In Sec. \ref{thermalization} we show that the collective spin observable quickly approaches its time-average value during time evolution. We also show that there is very good agreement between the diagonal ensemble average and the microcanonical average of the collective spin observable. In Sec. \ref{implementation} we discuss the physical implementation of the model. Finally, the conclusions are presented in Sec. \ref{s}.

\end{section}

\begin{section}{TMD Model}\label{TMD}

The TMD model describes a quantum system consisting of an ensemble of $N$ two-state atoms which interact with two boson modes via dipolar coupling. The Hamiltonian is given by ($\hbar=1$)
\begin{eqnarray}
&&\hat{H}=\hat{H}_{0}+\hat{H}_{\rm sb},\notag\\
&&\hat{H}_{0}=\omega_{a}\hat{a}^{\dag}\hat{a}+\omega_{b}\hat{b}^{\dag}\hat{b}+\Delta \hat{S}_{z},\notag\\
&&\hat{H}_{\rm sb}=\frac{2g_{a}}{\sqrt{N}}\hat{S}_{x}(\hat{a}^{\dag}+\hat{a})+\frac{2ig_{b}}{\sqrt{N}}\hat{S}_{y}(\hat{b}^{\dag}-\hat{b}).\label{model}
\end{eqnarray}
Here the first two terms in $\hat{H}_{0}$ contains the free boson terms where $\hat{a}^{\dag}$ and $\hat{b}^{\dag}$ are the creation operators corresponding to oscillators with frequencies $\omega_{a}$ and $\omega_{b}$. The third term in $\hat{H}_{0}$ describes the interaction between the collection of spins and the external applied magnetic field with strength $\Delta$. The interaction between the ensemble of spins and the two boson modes is given by $\hat{H}_{\rm sb}$ with $g_{a}$, $g_{b}$ being the coupling strengths and $\hat{S}_{\alpha}=\frac{1}{2}\sum_{l=1}^{N}\sigma_{l}^{\alpha}$ ($\alpha=x,y,z$) the collective spin operators, where $\sigma_{l}^{\alpha}$ is the Pauli operator for the $l$th spin.

Let us now discuss a few well-known limits of the model (\ref{model}). Setting ($g_{a}\neq 0$, $g_{b}=0$) or ($g_{a}= 0$, $g_{b}\neq0$) the Hamiltonian (\ref{model}) is equivalent to the Dicke model \cite{Emary2003}, which for $N=1$ reduces to the quantum Rabi model \cite{Larson2021}. In the limit of $N=1$ the TMD model reduces to the Jahn-Teller model which describes the dipolar interaction between a single spin and two vibrational modes \cite{Larson2008,Wang}.

\begin{figure}
\centering
\includegraphics[width=0.48\textwidth]{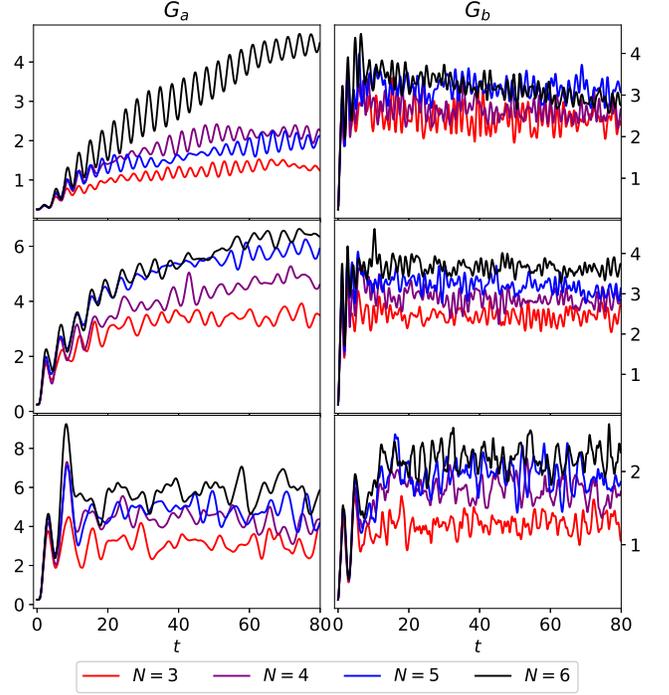}
\caption{Exact time evolution of the FOTOCs $G_a(t)$ (left) and $G_b(t)$ (right) for various $N$ from the initial state $|\psi_0 \rangle = \left|-S\right\rangle_x | 0 \rangle_a | 0 \rangle_b.$ The parameters are set to (top): $\omega_a = 1$, $\omega_b = 2$, $\Delta = 2$, $g_a = 0.5$, and $g_b = 3$. (middle): $\omega_a = 1$, $\omega_b = 2$, $\Delta = 2$, $g_a = 1.5$, and $g_b = 3$. (bottom): U(1)-symmetrical case with $\omega_a = \omega_b = 1$, $\Delta = 2$, and $g_a = g_b = 1.5$.}
\label{fig:fotoc}
\end{figure}


\end{section}

\begin{subsection}{Symmetries}
For general non-equal couplings $g_{a}\neq g_{b}$ the TMD model possesses a discrete $\mathbb{Z}_{2}$ symmetry, which is implemented by the parity operator, which is defined by
\begin{eqnarray}
&&\hat{\Pi}=\hat{\Pi}_{\rm spin}\otimes\hat{\Pi}_{\rm boson},\quad \hat{\Pi}_{\rm spin}=\sigma^{z}_{1}\otimes\ldots\otimes\sigma^{x}_{N},\notag\\
&&\hat{\Pi}_{\rm boson}=(-1)^{\hat{a}^{\dag}\hat{a}+\hat{b}^{\dag}\hat{b}},
\end{eqnarray}
where we have $\hat{\Pi}\hat{H}\hat{\Pi}=\hat{H}$. In the spacial case $\omega_{a}=\omega_{b}=\omega$ and $g_{a}=g_{b}=g$ the TMD model becomes U(1) invariant. Indeed, it is straightforward to show that the Hamiltonian (\ref{model}) commutes with the charge $\hat{C}=\hat{a}^{\dag}_{l}\hat{a}_{l}-\hat{a}^{\dag}_{r}\hat{a}_{r}+\hat{S}_{z}$, which is the symmetry group generator, where we have defined $\hat{a}_{l}=(\hat{a}-\hat{b})/\sqrt{2}$ and $\hat{a}_{r}=(\hat{a}+\hat{b})/\sqrt{2}$. Finally, the total Hilbert space is spanned in the basis $\{|m\rangle|n\rangle_a|n\rangle_b\}$ where $\hat{S}_{z}|m\rangle=m|m\rangle$ ($m=-S,\ldots,S$) with $S=N/2$ and $|n\rangle_{a(b)}$ is the Fock state of bosonic mode $a(b)$ with occupation number $n_{a(b)}$.

\begin{figure}[tp]
\centering
\includegraphics[width=0.48\textwidth]{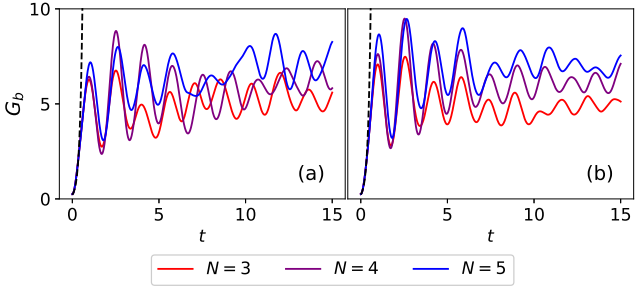}
\caption{Exact time evolution of the FOTOC $G_b (t)$ for (a) $g_b=4.5$, $g_a=0.5$ and (b) $g_b=4.5$, $g_a=1.5$ for various $N$ compared to the growth predicted by the largest classical Lyapunov exponent (dashed line).}
\label{fig:fotoc_expfit}
\end{figure}
\end{subsection}

\begin{subsection}{Quantum phase transition}
In the thermodynamic limit $S\rightarrow \infty$ the TMD model exhibits a quantum phase transition between a normal phase and a superradiant phase. For non-equal couplings $g_{a}\neq g_{b}$ the quantum phase transition is related to the spontaneous breaking of a $\mathbb{Z}_{2}$ symmetry at the critical coupling $g_{\rm c}=\sqrt{\omega\Delta}/2$, where for simplicity we have set $\omega_{a}=\omega_{b}=\omega$. The quantum phases of the system are a normal phase for $g_{a},g_{b}<g_{\rm c}$ characterized by zero mean-field bosonic excitations of both modes $\langle \hat{a}^{\dag}\hat{a}\rangle/S=\langle\hat{b}^{\dag}\hat{b}\rangle/S=0$ and mean collective spin orientation along the $z$-direction $\langle\hat{S}_{z}\rangle/S=-1$, and a superradiant phase, for which we have that  $\langle\hat{a}^{\dag}\hat{a}\rangle/S=2(g_{a}/\omega)^{2}(1-g^{4}_{\rm c}/g^{4}_{a})$, $\langle\hat{b}^{\dag}\hat{b}\rangle/S=0$, $\langle\hat{S}_{z}\rangle/S=-g_{\rm c}^{2}/g^{2}_{a}$ for $g_{a}>g_{\rm c}$, $g_{a}>g_{b}$, and respectively $\langle\hat{a}^{\dag}\hat{a}\rangle/S=0$, $\langle\hat{b}^{\dag}\hat{b}\rangle/S=2(g_{b}/\omega)^{2}(1-g^{4}_{\rm c}/g^{4}_{b})$, $\langle\hat{S}_{z}\rangle/S=-g_{\rm c}^{2}/g^{2}_{b}$ for $g_{b}>g_{\rm c}$, $g_{b}>g_{a}$ \cite{Ivanov2013,Fan}. We emphasize that due to the parity symmetry in the superradiant phase only one of the bosonic modes can be macroscopically excited. In the U(1)-symmetric case the superradiant phase for $g>g_{\rm c}$ is characterized by $\langle\hat{a}^{\dag}\hat{a}\rangle/S=2(g/\omega)^{2}(1-g^{4}_{\rm c}/g^{4})\cos^{2}(\phi)$, $\langle\hat{b}^{\dag}\hat{b}\rangle/S=2(g/\omega)^{2}(1-g^{4}_{\rm c}/g^{4})\sin^{2}(\phi)$ and $\langle\hat{S}_{z}\rangle/S=-g_{\rm c}^{2}/g^{2}$, where the phase $\phi$ remains undetermined which is a result of arbitrariness in the choice of a direction of spontaneous symmetry breaking \cite{Ivanov2013}, (see Appendix \ref{QPT}). In the symmetry-broken phase the energy spectrum of the TMD model contains one gapless Godstone mode and two gapped amplitude modes \cite{Ivanov2013,Porras2012,Ivanov2015}.

\end{subsection}

\begin{section}{Fidelity out-of-time-order correlators}\label{fotoc_sec}
We now discuss quantum information scrambling in the TMD model by investigating the behaviour of the fidelity out-of-time-order correlator in the quantum phases. 

The out-of-time-order correlation function
\begin{equation}
F(t)=\langle \hat{W}^{\dag}(t)\hat{V}^{\dag}\hat{W}(t)\hat{V}\rangle,\label{otoc}
\end{equation}
is a quantity which probes the spread of quantum information and signals the presence of quantum chaos through its exponential growth, characterized by a quantum Lyapunov exponent $\lambda_{\rm Q}$. Here, $\hat{W}$ and $\hat{V}$ are two initially ($t=0
$) commuting operators, where $\hat{W}(t)=e^{i \hat{H} t}\hat{W}e^{-i\hat{H} t}$ is an operator in the Heisenberg picture, and $\langle\cdot\rangle$ denotes the expectation value over the initial state $|\psi_{0}\rangle$. In the particular case when $\hat{W}=e^{i\delta\phi \hat{G}}$ with $\hat{G}$ being a Hermitian operator, $\delta\phi$ is a small perturbation, and $\hat{V}=|\psi_{0}\rangle\langle\psi_{0}|$ is the projection operator onto the initial state the expression (\ref{otoc}) is known as the fidelity out-of-time-order correlator (FOTOC) \cite{Schmitt2019,Swan2019}. As long as $\delta\phi\ll 1$ is a small perturbation we can expand the FOTOC $\mathcal{F}_{G}(t)=\langle \hat{W}^{\dag}_{G}(t)\hat{\rho}(0)\hat{W}_{G}(t)\hat{\rho}(0)\rangle$ in power series of $\delta\phi$, such that the dynamics of the FOTOC agrees with that of the variance of $\hat{G}$, namely
\begin{equation}
1-\mathcal{F}_{G}(t)=\delta\phi^{2}\left(\langle \hat{G}^{2}(t)\rangle-\langle \hat{G}(t)\rangle^{2}\right) =\delta\phi^{2}\Delta\hat{G}(t)^{2}.
\end{equation}\label{fotoc}

\begin{figure}[tp]
\centering
\includegraphics[width=0.48\textwidth]{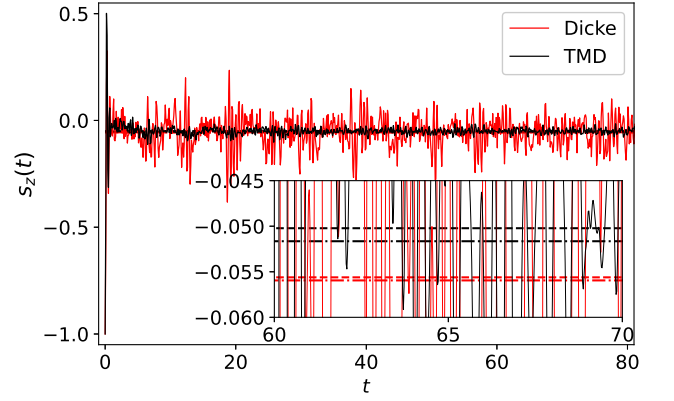}
\caption{Exact time evolution of the average value $s_{z}(t)=\langle\psi(t)|\hat{s}_{z}|\psi(t)\rangle$ of the collective spin operator $\hat{s}_{z}=\hat{S}_{z}/j$ (black line) for Hamiltonian (\ref{model}) and $N=4$. We set $\omega_a = 1$, $\omega_b = 2$, $\Delta = 2$, $g_a = 1.5$, and $g_{b}=3$. The initial state is $|\psi_0\rangle=\left|-S\right\rangle_z |5\rangle_a |10\rangle_b$. The long-time average $\langle\bar{s}_{z}\rangle$ is shown (black dashed line) with the ME prediction $\langle\bar{s}_{z}\rangle_{\rm ME}$ (black dash dot). Exact time evolution of $s_{z}(t)$ for the Dicke model for $g_{b}=0$ (red line). The initial state is $|\psi_0 \rangle =\left|-S\right\rangle_z|5\rangle_a$. The long-time average is shown with (red dashed line) along with the ME prediction (red dash dot line).}
\label{fig:Sz}
\end{figure}We discuss the behaviour of the FOTOC by setting $\hat{G}=(\hat{a}^{\dag}+\hat{a})/2=\hat{G}_{a}$ and $\hat{G}=(\hat{b}^{\dag}+\hat{b})/2=\hat{G}_{b}$, and respectively $G_{a}(t)=\Delta \hat{G}_{a}(t)^{2}$, $G_{b}(t)=\Delta \hat{G}_{b}(t)^{2}$.

In the normal phase $g_{a},g_{b}<g_{\rm c}$, both FOTOCs $G_{a}(t)$ and $G_{b}(t)$ oscillate with an amplitude independent on $N$. In the superradiant phase for $g_{a}<g_{\rm c}$ and $g_{b}>g_{\rm c}$ we observe initial exponential growth of $G_{b}(t)$ in the beginning of the evolution, which is associated with the macroscopically excited $b$-mode, and  slow non-exponential growth of $G_{a}(t)$, which corresponds to the macroscopically non-excited $a$-mode, see Fig. \ref{fig:fotoc}(top). Similar behaviour of the FOTOCs is also observed for $g_{a},g_{b}>g_{\rm c}$ and $g_{a}<g_{b}$, see Fig. \ref{fig:fotoc}(middle). In both cases $G_{b}(t)$ approaches saturation after the initial fast exponential growth, where it remains to oscillate with a small amplitude. The long-time behaviour of $G_{a}(t)$ is also characterized by saturation, see Fig. \ref{fig:fotoc}(middle). In the U(1)-symmetrical case and for $g_{a},g_{b}>g_{\rm c}$  we observe that both $G_{a}(t)$ and $G_{b}(t)$ show initial exponential growth and after that quickly reach a value of saturation, around which they oscillate for all subsequential times, see Fig. \ref{fig:fotoc}(bottom).

Recently, it was shown that the Lipkin-Meshkov-Glick and Dicke models may exhibit unstable points with equal classical and quantum Lyapunov exponents \cite{Cameo}. Thus, for an initial state centered at the unstable point, the FOTOC may show exponential behaviour. However, exponential growth, but with lower saturation and larger long-time oscillations, can still occur for initial states surrounding the unstable point, even though the classical Lyapunov exponent there is zero. Here we follow the method provided in \cite{Cameo} to characterize the behaviour of the FOTOC in the quantum phases of the TMD model. For $g_{a}<g_{\rm c}$ and $g_{b}>g_{\rm c}$ we find one positive Lyapunov exponent. This resembles the result for the Lipkin-Meshkov-Glick and Dicke models \cite{Cameo}.  Remarkably, for $g_{a},g_{b}>g_{\rm c}$ and $g_{a}<g_{b}$ the classical Lyapunov exponent associated with the unstable point is zero for the parameter regime that we use in Fig. \ref{fig:fotoc}. Additionally, we find that for the U(1)-symmetric TMD model the classical Lyapunov exponent is zero for all sets of parameters (see, Appendix \ref{UP} for more details). Hence in these cases the fast exponential growth of the FOTOC cannot be related to the classical unstable point, but is solely a property of the quantum dynamics of the system. In Fig. \ref{fig:fotoc_expfit} we plot the FOTOC for a different parameter regime, where there exist classical Lyapunov exponents that are non-zero, and find that indeed the growth of the FOTOC is governed by the largest classical Lyapunov exponent $ \lambda_{\rm Cl}$ as $G_b(t) \sim e^{\lambda_{\rm Q} t}$, where $\lambda_{\rm Q} = 2 \lambda_{\rm Cl}$.    

\begin{figure}[tp]
\centering
\includegraphics[width=0.48\textwidth]{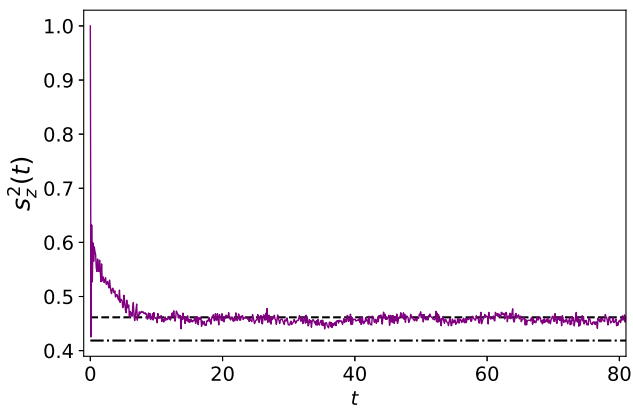}
\caption{Exact time evolution of $s^{2}_{z}(t)$ for Hamiltonian (\ref{model}) and $N = 4$ (purple solid line),  DE (dashed line) and ME (dashdot line) predictions. The initial state is $|\psi_0 \rangle = \left|-S\right\rangle_z|5\rangle_a|10\rangle_b$. The parameters are set to $\omega_a = 1$, $\omega_b = 2$, $\Delta = 2$, $g_a = 1.5$, and $g_b = 3$.}
\label{fig:Sz2}
\end{figure}

\end{section}

\begin{section}{Thermalization}\label{thermalization}
In the following we discuss thermalization of the collective spin observable. We show that the observable thermalizes in the quantum-chaotic superradiant phase.

A quantum observable thermalizes if its value agrees with the appropriately chosen statistical ensemble prediction at the corresponding initial energy for the most times of its evolution \cite{Alessio}. Consider that the system is prepared in the initial state $\left|\psi_{0}\right\rangle=\sum_{k}c_{k}|E_{k}\rangle$, where $\hat{H}|E_{k}\rangle=E_{k}|E_{k}\rangle$ and $c_{k}=\langle E_{k}|\psi_{0}\rangle$ which evolves under the system Hamiltonian as $\left|\psi(t)\right\rangle=e^{-i\hat{H} t}\left|\psi_{0}\right\rangle$. Then the evolution of an observable $\hat{O}$ is $O(t)=\langle\psi(t)|\hat{O}|\psi(t)\rangle$ and its long-time average is given by
\begin{equation}
    \langle \bar{O}\rangle=\lim_{\tau\rightarrow\infty}\frac{1}{\tau}\int_{0}^{\tau}O(t) dt=
    {\rm Tr}[\hat{\rho}_{\rm DE}\hat{O}],
\end{equation}
where $\hat{\rho}_{\rm DE}=\sum_{k}|c_{k}|^{2}|E_{k}\rangle\langle E_{k}|$ is the density matrix of the so-called diagonal ensemble (DE). The Eigenstate Thermalization Hypothesis (ETH) provides an explanation for the emergence of thermalization in an isolated quantum system \cite{Deutsch1991}. According to the ETH the expectation value of a thermalizing observable defined by $O(t)=\sum_{k}|c_{k}|^{2}|E_{k}\rangle\langle E_{k}|$ is equal to the microcanonical prediction at the corresponding energy. The microcanonical ensemble (ME) average of a observable $\hat{O}$ is given by
\begin{equation}
    \langle O\rangle_{\rm ME}(E_{0})={\rm Tr}[\hat{\rho}_{\rm ME}\hat{O}]=\frac{1}{\mathcal{N}}
    \sum_{m:|E_{m}-E_{0}|<\delta E}O_{mm},
\end{equation}
where $E_{0}=\langle\psi_{0}|\hat{H}|\psi_{0}\rangle$ is the energy of the system, $O_{mm}=\langle E_{m}|\hat{O}|E_{m}\rangle$, and $\mathcal{N}$ is the number of eigenstates of $\hat{H}$ that are inside an energy shell of width $2\delta E$ around the energy $E_{0}$. Finally, we define the long-time average of the temporal fluctuations of the expectation value of the observable $\hat{O}$ as
\begin{equation}
    \delta_{O}^{2}=\lim_{\tau\rightarrow\infty}\frac{1}{\tau}\int_{0}^{\tau}O(t)^{2}dt-\bar{O}^{2}.\label{LTF}
\end{equation}
Thermalization requires that after equilibration to a microcanonical state the long-time fluctuations of an observable $\hat{O}$ (\ref{LTF}) around this state are small \cite{Nation2019}. More precisely, the fluctuations should be bounded in terms of some effective dimension of the state of the system.

\begin{figure}[tp]
\centering
\includegraphics[width=0.48\textwidth]{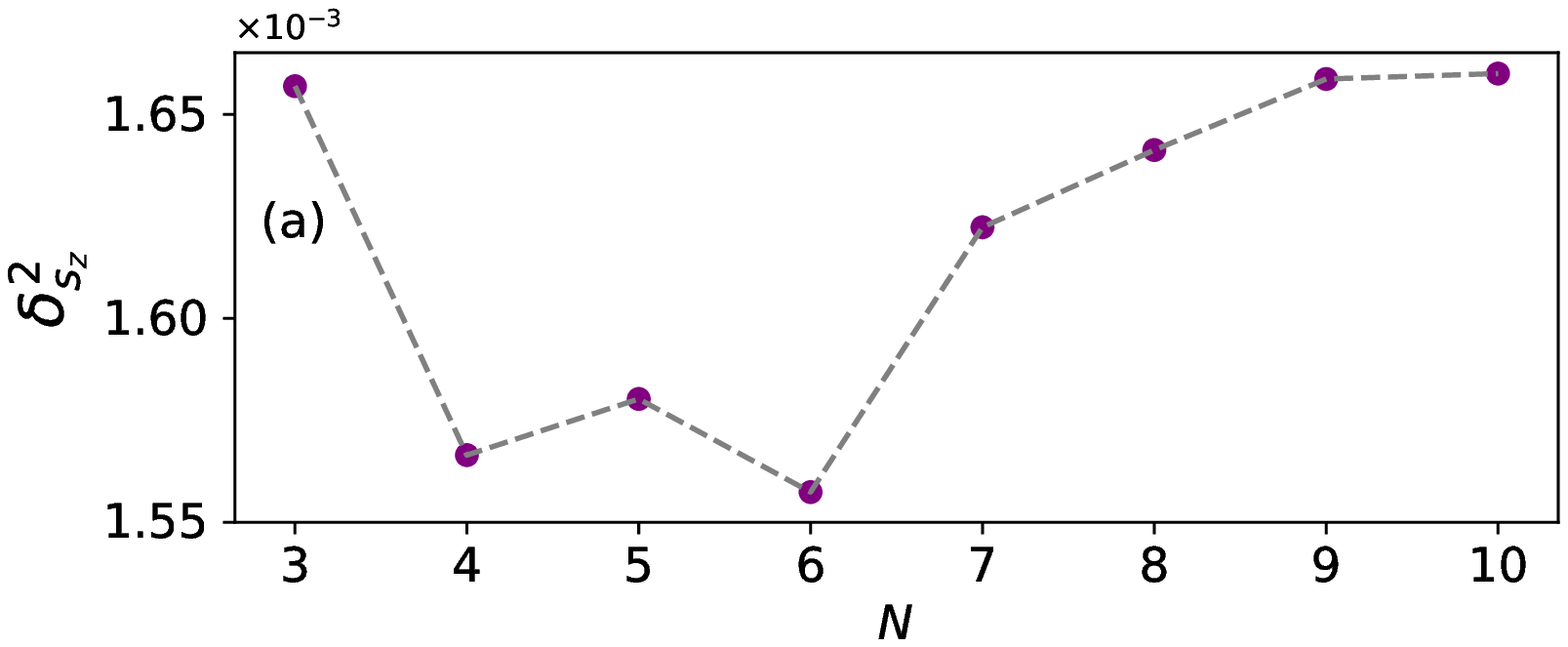}
 \includegraphics[width=0.48\textwidth]{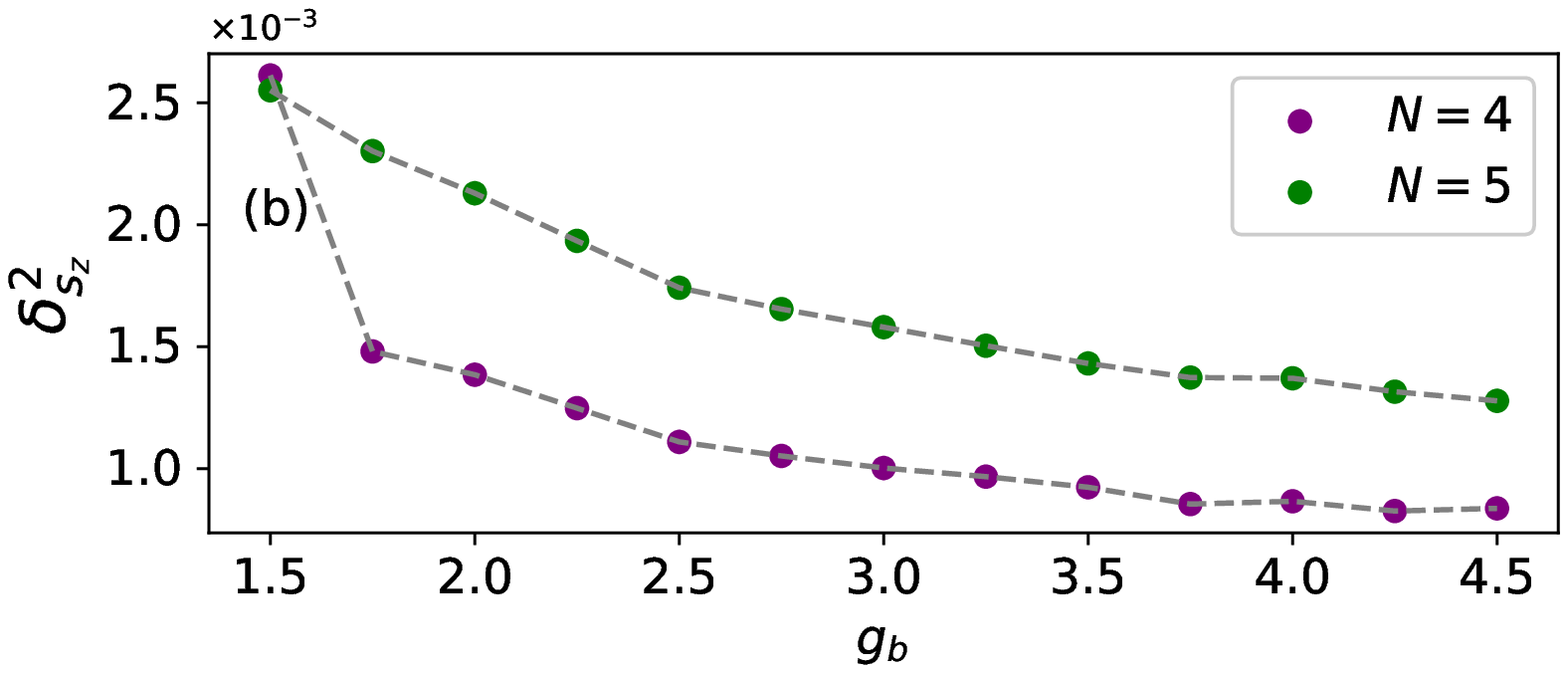}
  \includegraphics[width=0.48\textwidth]{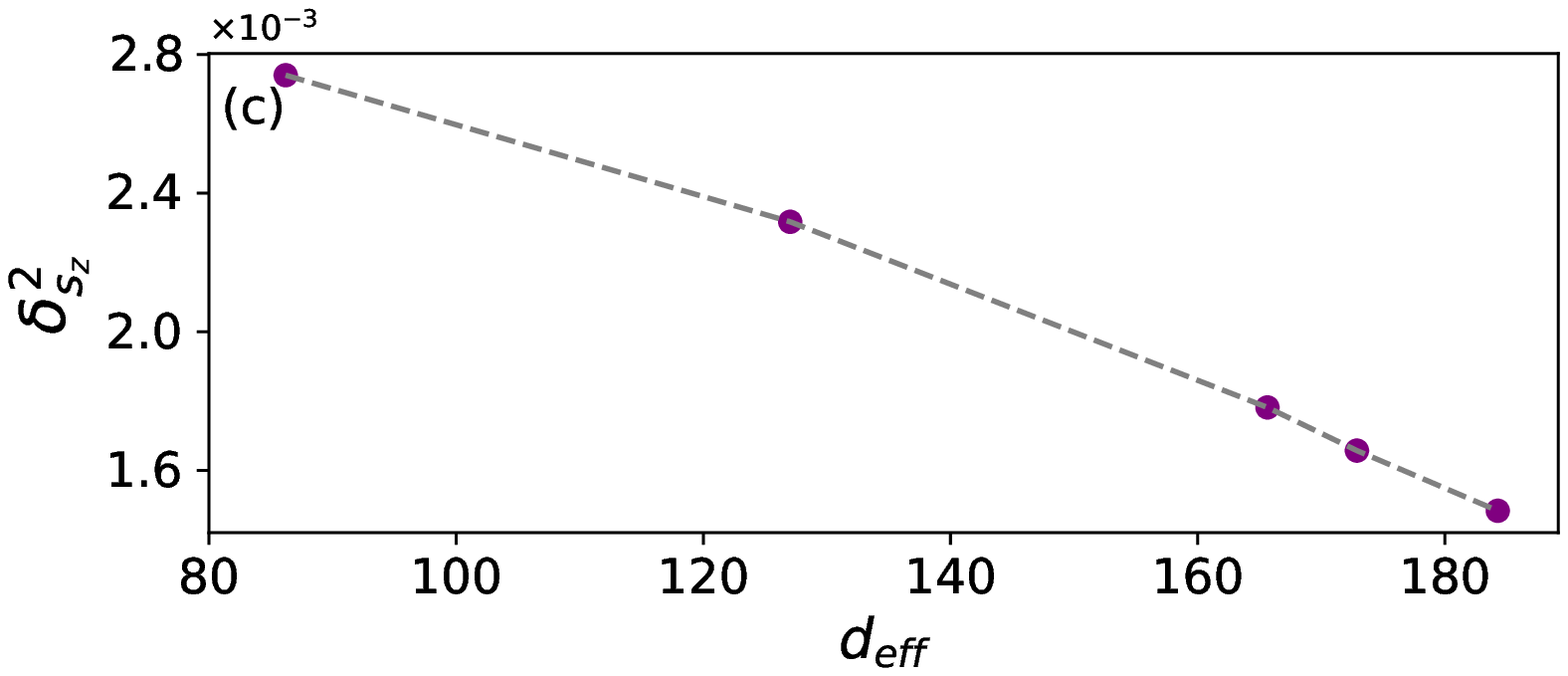}
\caption{Long-time average of the temporal fluctuation $\delta^{2}_{s_{z}}$, (a) as a function of $N$ for $g_{b}=3$, (b) as a function of the spin-boson coupling $g_{b}$ for $N=4$. The other parameters are set to $\omega_{a}=1$, $\omega_{b}=2$, $\Delta=2$, $g_{a}=1.5$, and the initial state is $|\psi_0\rangle=\left|-S\right\rangle_z |5\rangle_a |10\rangle_b$.
(c) $\delta^{2}_{s_{z}}$ as a function of $d_{\rm eff}$ for $N = 3$, obtained through varying $g_b$ in the Hamiltonian from $1.5$ to $3.5$ in increments of $0.5$.}
\label{fig:Szfluct}
\end{figure}

In Fig. \ref{fig:Sz} we show the evolution of the expectation value of the collective spin operator $\hat{s}_{z}=\hat{S}_{z}/S$. We observe that after the initial quench the observable very quickly approaches saturation with suppressed temporal fluctuations around the long-time average expectation value $\langle\bar{s}_{z}\rangle$. Furthermore, we compare $\langle\bar{s}_{z}\rangle$ with $\langle s_{z}\rangle_{\rm ME}(E_{0})$, where the microcanonical shell is chosen such that the ME prediction gives nearly the same result despite small fluctuations around the value of $\delta E$. This is in accordance with the ETH statement that the microcanonical prediction is identical to the expectation value in any eigenstate within the energy shell. We see that TMD model shows nearly perfect agreement with the ME prediction even for a very small number of spins. For comparison we also plot the evolution of $\langle\bar{s}_{z}\rangle$ for the Dicke model by setting $g_{a}=0$. Although the long-time average is in good agreement with the ME prediction, the temporal oscillations are with higher amplitudes in comparison to the TMD model.

In Fig. \ref{fig:Sz2} we show the evolution of the expectation value of the two-body observable $s_{z}^{2}(t)=\langle\psi(t)|\hat{s}^{2}_{z}|\psi(t)\rangle$ with $\hat{s}^{2}_{z}=\hat{S}^{2}_{z}/S^{2}$ for $N=4$ spins. We see that $s_{z}^{2}(t)$ equilibrates quickly to its long-time average value $\langle\bar{s}_{z}^{2}\rangle$. However, for this small number of spins, $\langle\bar{s}_{z}^{2}\rangle$ deviates from the ME prediction $\langle s^{2}_{z}\rangle_{\rm ME}(E_{0})$. Note that the exact scaling of $\langle s^{2}_{z}\rangle_{\rm ME}(E_{0})$ with $N$ is impossible to calculate with good enough precision due to the computational resources needed for an increasingly larger size of truncated total Hilbert space.

In Fig. (\ref{fig:Szfluct}) is shown the long-time average fluctuations of the observable $\hat{s}_{z}$. We see that even for a small number of spins $N$, the fluctuations $\delta_{s_{z}}^{2}$ are suppressed. Moreover, increasing the spin-boson coupling $g_{b}$ causes the fluctuations $\delta_{s_{z}}^{2}$ to gradually decrease. Let us now define the effective system size as $d_{\rm eff}=(\sum_{k}|c_{k}|^{4})^{-1}$ where $c_{k}=\langle E_{k}|\psi_0\rangle$. Thermalization requires that $d_{\rm eff}\gg 1$, which guarantees that the initial state is composed from a large number of eigenstates $|E_{k}\rangle$. Thus, the bosonic degrees of freedom can be considered as an effective thermal bath coupled to the collective spin system. The ETH ansatz also requires that after equilibration to a microcanonical state the fluctuations $\delta_{s_{z}}^{2}$ around this state decrease with the effective system size $d_{\rm eff}$. In Fig. (\ref{fig:Szfluct})c we plot this relation, obtained through varying the coupling $g_{b}$. We see that $\delta_{s_{z}}^{2}$ decreases with $d_{\rm eff}$, which leads to thermalization of the collective spin observable.

\end{section}

\begin{section}{Physical Implementation with Trapped Ions}\label{implementation}
An experimental observation of the predicted quantum chaos and thermalization of the collective spin observable in the TMD model can be achieved in a system of trapped ions \cite{Wineland1998,Schneider2012}. The two common center-of-mass modes along the transverse directions are the oscillators in the TMD model. We assume that all other vibrational modes are far separated in frequency and thus can be neglected. Each ion has two metastable levels $\left|\uparrow\right\rangle$, $\left|\downarrow\right\rangle$ with transition frequency $\omega_{0}$ which form an effective spin system. The spin-phonon interactions can be created by a direct two-photon optical transition as in the case of optical levels, or alternatively one can use radio-frequency, or hyperfine levels, where the spin-phonon couplings are driven by a Raman-type interaction. We consider that in each spatial direction two lasers with frequencies $\omega_{{\rm b},\alpha}=\omega_{0}-\Delta+(\omega_{\alpha}-\delta_{\alpha})$,  $\omega_{{\rm r},\alpha}=\omega_{0}-\Delta-(\omega_{\alpha}-\delta_{\alpha})$ drive simultaneously the blue- and the red-sideband transitions of the center-of-mass vibrational modes. Here $\omega_{\alpha}$ ($\alpha=x,y$) are the transverse trap frequencies, $\Delta$ and $\delta_{\alpha}$ are small detunings, $\delta_{\alpha},\Delta\ \ll \omega_{0},\omega_{\alpha}$. After moving to the interaction picture with respect to the spin and oscillator dynamics and after performing a rotating wave approximation (RWA) the interaction Hamiltonian in the Lamb-Dicke limit is
\begin{eqnarray}
\hat{H}_{I}&=&\frac{1}{\sqrt{N}}\sum_{l=1}^{N}\{\eta_{x}\Omega_{x}\sigma^{+}_{l}e^{i\Delta t} (\hat{a}^{\dag}_{x}e^{i\delta_{x}t}+\hat{a}_{x}e^{-i\delta_{x}t})\notag\\
&&+\eta_{y}\Omega_{y}\sigma^{+}_{l}e^{i\Delta t}(\hat{a}^{\dag}_{y}e^{i\delta_{y}t}-\hat{a}_{y}e^{-i\delta_{y}t})+{\rm H.c.}\},\label{HII}
\end{eqnarray}
where $\Omega_{\alpha}$ are the peak Rabi frequencies and $\eta_{\alpha}$ are the Lamb-Dicke parameters. In the rotating frame where the interaction Hamiltonian (\ref{HII}) is time-independent, $\hat{H}_{I}$ is identical to (\ref{model}) with $\omega_{a}=\delta_{x}$, $\omega_{b}=\delta_{y}$,
$g_{a}=\eta_{x}\Omega_{x}$, $g_{b}=\eta_{y}\Omega_{y}$, and respectively $\hat{a}_{x}=\hat{a}$, $\hat{a}_{y}=\hat{b}$.

The measurement of the FOTOC can be achieved by observing the imperfect time reversal dynamics, in which after the initial state preparation (i) the system unitarily evolves under the action of Hamiltonian (\ref{model}), (ii) a perturbation $\hat{W}=e^{i\delta\phi \hat{G}}$ is applied, (iii) followed by time-reversed dynamics, and finally (iv) a projective measurement of the final overlap with the initial state is performed \cite{Swan2020,Garttner2018}. Such time-reversal dynamics used to measure FOTOC was experimentally demonstrated in a long-range Ising model using ions in a Penning trap \cite{Garttner2017}. Let us now discuss the possible set of parameters to realize quantum thermalization. Consider for example $\omega_{a}/2\pi=20$ kHz, $\omega_{b}/2\pi=\Delta/2\pi=40$ kHz, and $g_{a}/2\pi=30$ kHz, $g_{b}/2\pi=60$ kHz, see Fig. (\ref{fig:Sz}). Thermalization occurs after an interaction time of order of $t\approx 100$ $\mu$s, which is shorter than the experimentally measured coherence time in typical trapped ion setups. Finally, the collective spin observable can be detected by laser-induced fluorescence, which is imaged on a CCD camera.

\end{section}

\begin{section}{Summary}\label{s}
We have discussed the onset of chaos and thermalization in the TMD model. Using the FOTOC as a measure of quantum information scrambling, we have studied the quantum phases of the TMD model. For the $\mathbb{Z}_{2}$-symmetric TMD model the superradiant phase is characterized by exponential growth of the FOTOC in the beginning of the time evolution for the macroscopically excited bosonic mode, whereas the FOTOC corresponding to the non-excited bosonic mode displays slow non-exponential dynamics. For the U(1)-symmetric TMD mode we have shown that both FOTOCs are characterized by initial exponential growth. We have found that the exponential growth cannot always be associated with an unstable point, since for some sets of parameters the classical Lyapunov exponent is zero.

Furthermore, we have investigated thermalization of the collective spin observable in the TMD model. We have shown that the collective spin observable quickly approaches its long-time average value and that the diagonal ensemble average agrees with the microcanonical average of the observable even for a small number of spins. We also have shown that the long-time average of its fluctuations is small and decreases when the spin-phonon coupling and effective system size are increased. Additionally, we discuss thermalization of the two-body spin observable. Although fluctuations around its the long-time average value are small, there is no good agreement with the microcanonical prediction for a small number of spins.

Our model can be implemented with trapped ions, where the collective spin is formed by the two metastable levels of the ions and the two bosons are the collective center-of-mass vibrational modes. We estimate that thermalization of the spin observable would occur at a much shorter time compared to the decoherence time in an ion trap. This allows for the study of onset of quantum chaos and ergodicity in a closed system. 

\section*{Acknowledgments}

A. V. K. and P. A. I. acknowledges financial support by the Sofia University Grant No 80-10-61.

\end{section}

\appendix
\begin{section}{Quantum Phase Transition}\label{QPT}
For the sake of the reader's convenience we discuss the onset of quantum phase transition in the TMD model.
We introduce a Holstein-Primakoff transformation $\hat{S}_{+}=\hat{c}^{\dag}\sqrt{2S-\hat{c}^{\dag}\hat{c}}$, $\hat{S}_{z}=\hat{c}^{\dag}\hat{c}-S$, which maps the collective spin excitation onto a bosonic excitation. Then in the thermodynamic limit $S\rightarrow\infty$ the TMD Hamiltonian becomes
\begin{equation}
\hat{H}=\sum_{l=1}^{3}\frac{\hat{p}^2_{l}}{2}+\frac{1}{2}\sum_{k,l=1}^3B_{kl}\hat{x}_l \hat{x}_k-e_0,
\end{equation}
where $e_0=\frac{1}{2}(\omega_a+\omega_b+\Delta)-S\Delta$ and $\hat{x}_{l}$, $\hat{p}_{l}$ ($l=1,2,3$) are the position and momentum operators, which are related to the bosonic operators via
\begin{eqnarray}
&&\hat{a}=\sqrt{\frac{\omega_a}{2}}\hat{x}_{1}+\frac{i}{\sqrt{2\omega_{a}}}\hat{p}_{1},\notag\\
&&\hat{b}=\frac{1}{2}\{\sqrt{\omega_{b}}\left(\frac{\hat{x}_2}{\sqrt{m_2}}+\frac{\hat{x}_3}{\sqrt{m_3}}\right)+\frac{i}{\sqrt{\omega_b}}(\sqrt{m_2}\hat{p}_2+\sqrt{m_3}\hat{p}_3)\},\notag\\
&&\hat{c}=\frac{1}{2}\{\sqrt{\Delta}\left(\frac{\hat{x}_2}{\sqrt{m_2}}-\frac{\hat{x}_3}{\sqrt{m_3}}\right)+\frac{i}{\sqrt{\Delta}}(\sqrt{m_2}\hat{p}_2-\sqrt{m_3}\hat{p}_3)\}.
\end{eqnarray}
with $m_2^{-1}=\left(1-\frac{2g_b}{\sqrt{\omega_b\Delta}}\right)$, $m_3^{-1}=\left(1+\frac{2g_b}{\sqrt{\omega_b\Delta}}\right)$. The $3\times3$ real and symmetric matrix $B_{kl}$ is given by
\begin{equation}
B_{kl} = \begin{bmatrix}
\omega_a^2 & g_a\sqrt{\frac{2\omega_a\Delta}{m_2}} & -g_a\sqrt{\frac{2\omega_a\Delta}{m_3}} \\
g_a\sqrt{\frac{2\omega_a\Delta}{m_2}} & \frac{\epsilon^2}{m_2}& \frac{\lambda}{\sqrt{m_2m_3}}\\
-g_a\sqrt{\frac{2\omega_a\Delta}{m_3}} & \frac{\lambda}{\sqrt{m_2m_3}} & \frac{\epsilon^2}{m_3}\\
\end{bmatrix}, \label{matrix}
\end{equation}
where $\epsilon^2=(\omega^2_b+\Delta)/2$ and $\lambda=(\omega^2_b-\Delta)/2$. The three collective spin-boson modes are given by the eigenvalues of (\ref{matrix}) which are real and positive as long as $g_a,g_b\le g_{\rm c}$ with $g_{\rm c}=\sqrt{\omega\Delta}/2$. Therefore, in the thermodynamical limit the normal phase is characterized with vanishing density of bosonic excitations $\langle \hat{a}^{\dag}\hat{a}\rangle/S=\langle \hat{b}^{\dag}\hat{b}\rangle/S=0$ and collective spin pointing along the $z$ axis $\langle \hat{S}_{z}\rangle/S=-1$.

Next, we displace each of the bosonic modes $\hat{a}\rightarrow \hat{a}+\sqrt{\alpha_a}$, $\hat{b}\rightarrow \hat{b}+\sqrt{\alpha_b}$, and $\hat{c}\rightarrow \hat{c}-\sqrt{\alpha_c}$, where $\alpha_a$, $\alpha_b$, and $\alpha_c$ are generally complex parameters in order of $S$. Substituting this transformation in the TMD Hamiltonian, the displacement parameters are determined by the condition that all linear terms in the bosonic operators are cancelled. We obtain
\begin{eqnarray}
&&\sqrt{\alpha_a}=\frac{2g_a}{\omega_a}\sqrt{\frac{k}{N}}r\cos\phi,\quad \sqrt{\alpha_b}=-i\frac{2g_b}{\omega_b}\sqrt{\frac{k}{N}}r\sin\phi,\notag\\
&&\frac{g^2_a\cos\phi}{g^2_{\rm c}N}\{k-r^2e^{i\phi}\cos\phi \}+\frac{g^2_b\sin\phi}{g^2_{\rm c}N}\{ik-r^2e^{i\phi}\sin\phi \}\notag\\
&&-e^{i\phi}=0,
\end{eqnarray}
where $\sqrt{\alpha_c}=re^{i\phi}$ and $k=N-r^2$. For $\mathbb{Z}_{2}$-symmetric case the solution is $\sqrt{\alpha_a}=\pm(g_a/\omega)\sqrt{2S(1-g_{\rm c}^4/g_a^4)}$, $\sqrt{\alpha_b}=0$, and $\sqrt{\alpha_c}=\pm\sqrt{S(1-g^2_{\rm c}/g_a^2)}$ for $g_a>g_{\rm c}$, $g_b<g_a$, and respectively $\sqrt{\alpha_a}=0$, $\sqrt{\alpha_b}=\pm i(g_b/\omega)\sqrt{2S(1-g_{\rm c}^4/g_b^4)}$, and $\sqrt{\alpha_c}=\pm i\sqrt{S(1-g^2_{\rm c}/g_b^2)}$ for $g_b>g_{\rm c}$, $g_a<g_b$. For continuous U(1) symmetry we have $|\alpha_a|+|\alpha_b|=(g/\omega)\sqrt{2S(1-g_{\rm c}^4/g^4)}$ and $\sqrt{|\alpha_c|}=\sqrt{S(1-g^2_{\rm c}/g^2)}$ for $g>g_{\rm c}$.

\end{section}

\begin{figure}
\centering
\includegraphics[width=0.45\textwidth]{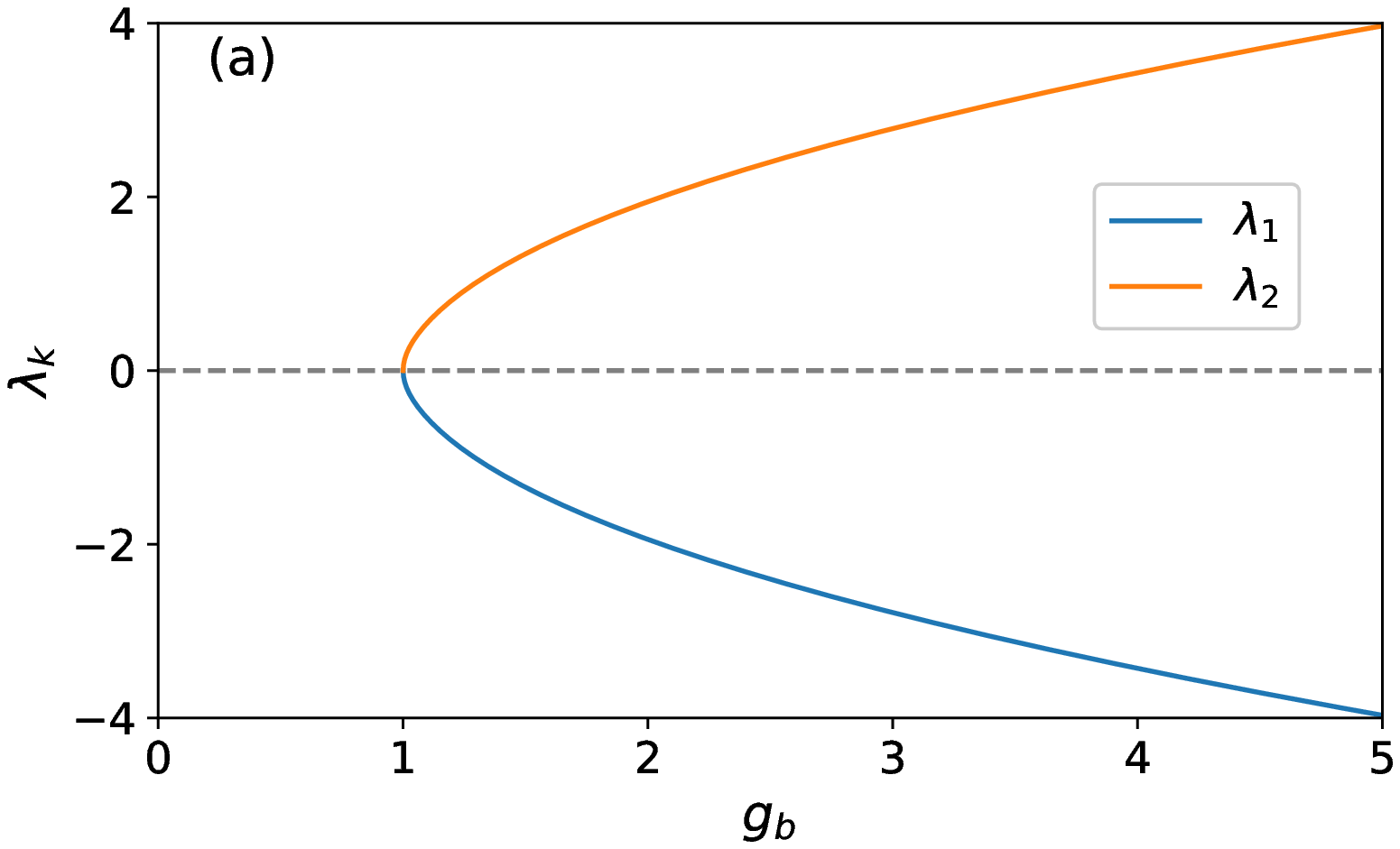}
\includegraphics[width=0.45\textwidth]{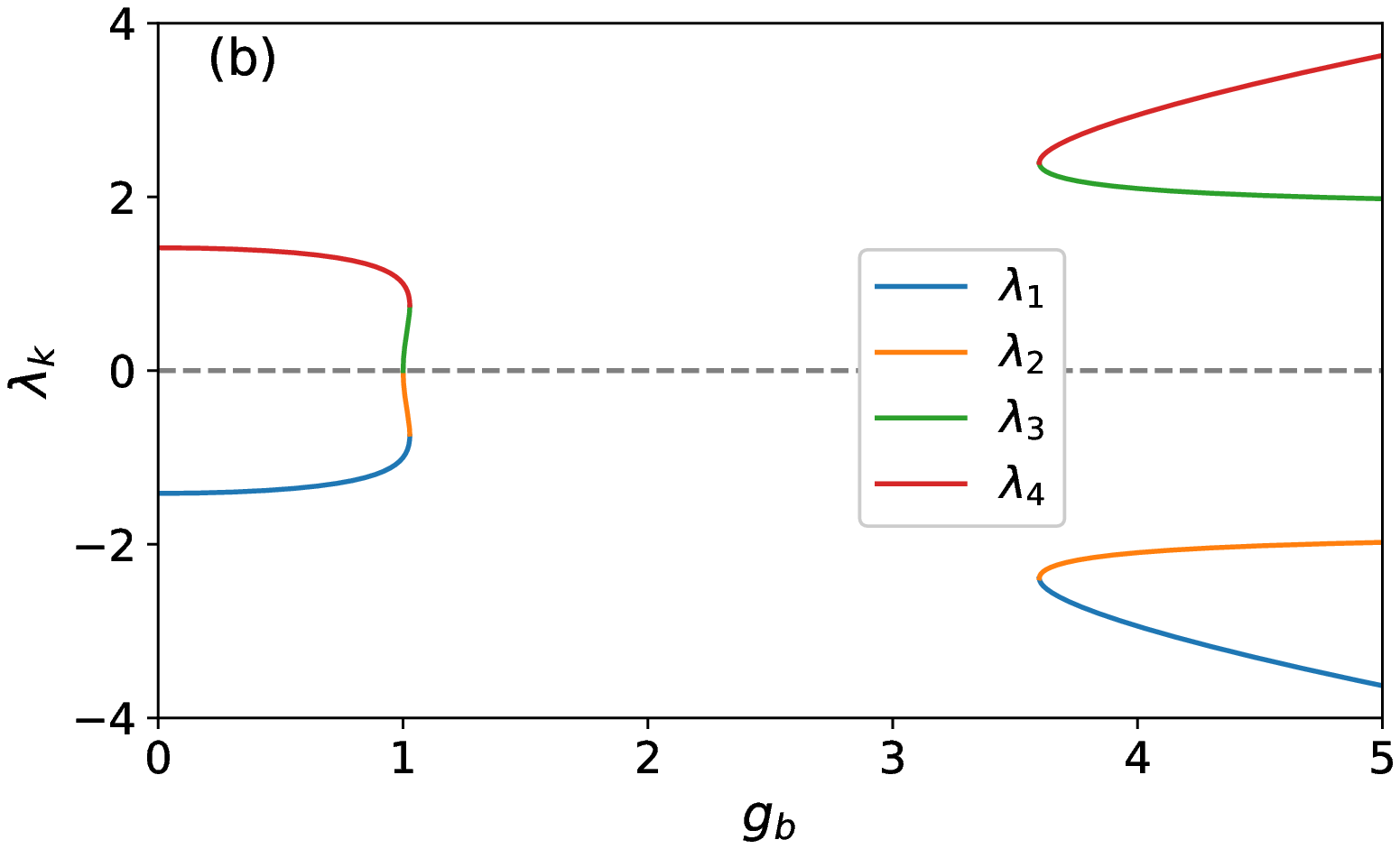}
\caption{Eigenvalues of Jacobian matrix (\ref{matrixJT}) as a function the coupling $g_{b}$. (a) The parameters are $\omega_{a}=1$, $\omega_{b}=2$, $\Delta=2$, and $g_{a}=0.5$. (b) The same but now $g_{a}=1.5$.}
\label{1}
\end{figure}
\begin{section}{Unstable points for TMD model}\label{UP}
We follow the method shown in \cite{Cameo} to calculate the unstable points that give rise to classical Lyapunov exponents. The mean-field Hamiltonian is obtained by taking the expectation value of $\hat{H}_{\rm TMD}$ on the mean-field ansatz,
\begin{equation}
|\psi_{\rm MF}\rangle=\prod_{l}|\theta_{l},\phi_{l}\rangle \otimes e^{\alpha \hat{a}^{\dag}-\alpha^{*}\hat{a}}
e^{\beta \hat{b}^{\dag}-\beta^{*}\hat{b}}|0_{a}\rangle|0_{b}\rangle.
\end{equation}
Here $|\theta_{l},\phi_{l}\rangle=\cos\left(\frac{\theta}{2}\right)\left|\downarrow_{l}\right\rangle+e^{i\phi}\sin\left(\frac{\theta}{2}\right)\left|\uparrow_{l}\right\rangle$ is the coherent spin state with $\theta=2\tan^{-1}\sqrt{\frac{Q^{2}+P^{2}}{4-(Q^{2}+P^{2})}}$ and $\phi=-\tan^{-1}\left(\frac{P}{Q}\right)$. The displacement parameters are $\alpha=\sqrt{\frac{S}{2}}(q_{a}+i p_{a})$ and $\beta=\sqrt{\frac{S}{2}}(q_{b}+i p_{b})$. The classical Hamiltonian is obtained in the limit $\mathcal{H}_{\rm TMD}=\lim_{S\rightarrow\infty}\langle \psi_{\rm MF}|\hat{H}_{\rm TMD}|\psi_{\rm MF}\rangle/S$ such that
\begin{eqnarray}
\mathcal{H}_{\rm TMD}&=&\frac{\omega_{a}}{2}(q^{2}_{a}+p_{a}^{2})+\frac{\omega_{b}}{2}(q^{2}_{b}+p_{b}^{2})+\frac{\Delta}{2}(Q^{2}+P^{2})\notag\\
&&+2g_{a}\sqrt{1-\frac{Q^{2}+P^{2}}{4}}q_{a}Q+2g_{b}\sqrt{1-\frac{Q^{2}+P^{2}}{4}}\notag\\
&&\times p_{b}P-\frac{\Delta}{2}.\nonumber
\end{eqnarray}
Here the variables $Q$, $q_{a}$, $q_{b}$ are the generalized coordinates and respectively  $P$, $p_{a}$, $p_{b}$ are the generalized momenta. The classical Hamilton equations $d \mathbf{X}/dt=\mathbf{F}(\mathbf{X})$ for $\mathbf{X}=(Q,q_{a},q_{b},P,p_{a},p_{b})$ have a stationary point at $\mathbf{X}_{0}=(0,0,0,0,0,0)$, where $\mathbf{F}(\mathbf{X}_{0})=0$. The corresponding Jacobian matrix of $\mathbf{F}$ evaluated at $\mathbf{X}_{0}$ is given by
\begin{equation}
A = \begin{bmatrix}
0 & 0 & 0 & \Delta &0 &2g_{b} \\
0 & 0& 0 & 0 &\omega_{a} &0\\
0 & 0 & 0  & 2g_{b} & 0&\omega_{b}\\
-\Delta&-2g_{a}& 0  & 0 & 0&0\\
-2g_{a}&-\omega_{a}&0  & 0& 0 &0\\
0&0&-\omega_{b} & 0 &   0  &  0\\
\end{bmatrix}  . \label{matrixJT}
\end{equation}
In Fig. \ref{1} we plot the eigenvalues of the Jacobian matrix (\ref{matrixJT}). We see in Fig. \ref{1}(a) that for $g_{b}>g_{\rm c}$ and $g_{a}<g_{\rm c}$ there is only one possitive Lyapunov exponent. This case resembles the behaviour of the Dicke model found in \cite{Cameo}. However, for $g_{b}>g_{\rm c}$ but now $g_{a}>g_{\rm c}$ we find that there is a region of values $g_{b}$ where all eigenvalues of $A$ are complex, see Fig. \ref{1}(b). Therefore the exponential growth of the FOTOC for this case is not associated with the unstable point. We also find that for U(1)-symmetric case all eigenvalues are complex for all set of parameters.

\end{section}


\begin{thebibliography}{99}

\bibitem{Rigol2008} M. Rigol, V. Dunjko, and M. Olshanii, Nature (London) \textbf{452}, 854 (2008).

\bibitem{Eisert2014} J. Eisert, M. Friesdorf, and C. Gogolin, Nat. Phys. \textbf{11}, 124 (2014).

\bibitem{Alessio} L. D'Alessio, Y. Kafri, A. Polkovnikov, and M. Rigol, Adv. Phys. \textbf{65}, 239 (2016).

\bibitem{Gogolin2016} C. Gogolin and J. Eisert, Rep. Prog. Phys. \textbf{79}, 056001 (2016).

\bibitem{Neill2016} C. Neill \emph{et al}., Nat. Phys. \textbf{12}, 1037 (2016).

\bibitem{Smith2016} J. Smith, A. Lee, P. Richerme, B. Neyenhuis, P. W. Hess, P. Hauke, M. Heyl, D. A. Huse, and C. Monroe, Nat. Phys. \textbf{12}, 907 (2016).

\bibitem{Linden2009} N. Linden, S. Popescu, A. J. Short, and A. Winter, Phys. Rev. E \textbf{79}, 061103 (2009).

\bibitem{Altland2012} A. Altland and F. Haake, Phys. Rev. Lett. \textbf{108}, 073601 (2012).

\bibitem{Swan2019} R. J. Lewis-Swan, A. Safavi-Naini, J. J. Bollinger, and A. M. Rey, Nat. Commun. \textbf{10}, 1581 (2019).

\bibitem{Kaufman2016} A. M. Kaufman, M. E. Tai, A. Lukin, M. Rispoli, R. Schittko, P. M. Preiss, and M. Greiner, Science \textbf{353}, 794 (2016).

\bibitem{Clos2016} G. Clos, D. Porras, U. Warring, and T. Schaetz, Phys. Rev. Lett. \textbf{117}, 170401 (2016).

\bibitem{Maldacena2016} J. Maldacena, S. H. Shenker, and D. Stanford, J. High Energy Phys. \textbf{2016}, 106 (2016).

\bibitem{Swingle2018} B. Swingle, Nat. Phys. \textbf{14}, 988 (2018).

\bibitem{Mata2018} I. Carcia-Mata, M. Saraceno, R. A. Jalabert, A. J. Roncaglia, and D. A. Wisniacki, Phys. Rev. Lett. \textbf{121}, 210601 (2018).

\bibitem{Shenker2014} S. H. Shenker and D. Stanford, J. High Energy Phys. \textbf{2014}, 67 (2014).

\bibitem{Sekino} Y. Sekino and L. Susskind, J. High Energy Phys. \textbf{2008}, 065 (2008).

\bibitem{Markovic2022} D. Markovi\'c and M. \v{C}ubrovi\'c, J. High Energy Phys. \textbf{2022}, 23 (2022).

\bibitem{Heyl2018} M. Heyl, F. Pollmann, and B. D\'ora, Phys. Rev. Lett. \textbf{121}, 016801 (2018).

\bibitem{Shen2017} H. Shen, P. Zhang, R. Fan, and H. Zhai, Phys. Rev. B \textbf{96}, 054503 (2017).

\bibitem{Swan2020} R. J. Lewis-Swan, S. R. Muleady, and A. M. Rey, Phys. Rev. Lett. \textbf{125}, 240605 (2020).

\bibitem{Dag2019} C. B. Dag, K. Sun, and L.-M. Duan, Phys. Rev. Lett. \textbf{123}, 140602 (2019).

\bibitem{carlosprl2019} J. Ch\'avez-Carlos \emph{et al}., Phys. Rev. Lett. \textbf{122}, 024101 (2019).

\bibitem{Sun2020} Z.-H. Sun, J.-Q. Cai, Q.-C. Tang, Y. Hu, and H. Fan, Ann. Phys. (Berlin, Ger.) \textbf{532}, 1900270 (2020).

\bibitem{Kirkova2022} A. V. Kirkova, D. Porras, and P. A. Ivanov, Phys. Rev. A \textbf{105}, 032444 (2022).

\bibitem{Li2017} J. Li, R. Fan, H. Wang, B. Ye, B. Zeng, H. Zhai, X. Peng, and J. Du, Phys. Rev. X \textbf{7}, 031011 (2017).

\bibitem{Braumuller2022} J. Braum\"uller \emph{et al}., Nat. Phys. \textbf{18}, 172 (2022

\bibitem{Garttner2017} M. G\"arttner, J. G. Bohnet, A. Safavi-Naini, M. L. Wall, J. J. Bollinger, and A. M. Rey, Nat. Phys. \textbf{13}, 781 (2017).

\bibitem{Landsman2019} K. A. Landsman, C. Figgatt, T. Schuster, N. M. Linke, B. Yoshida, N. Y. Yao, and C. Monroe, Nature \textbf{567}, 61 (2019).

\bibitem{Joshi2020} M. K. Joshi, A. Elben, B. Vermersch, T. Brydges, C. Maier, P. Zoller, R. Blatt, and C. F. Roos, Phys. Rev. Lett. \textbf{124}, 240505 (2020).

\bibitem{Green2022} A. M. Green \emph{et al}., Phys. Rev. Lett. \textbf{128}, 140601 (2022).

\bibitem{Pegahan2021} S. Pegahan, I. Arakelyan, and J. E. Thomas, Phys. Rev. Lett. \textbf{126}, 070601 (2021).

\bibitem{Ivanov2013} P. A. Ivanov, D. Porras, S. S. Ivanov, and F. Schmidt-Kaler, J. Phys. B: At. Mol. Opt. Phys. \textbf{46}, 104003 (2013).

\bibitem{Fan} J. Fan, Z. Yang, Y. Zhang, J. Ma, G. Chen, and S. Jia, Phys. Rev. A \textbf{89}, 023812 (2014).

\bibitem{Porras2012} D. Porras, P. A. Ivanov, and F. Schmidt-Kaler, Phys. Rev. Lett. \textbf{108}, 235701 (2012).

\bibitem{Ivanov2015} P. A. Ivanov, J. Low. Temp. Phys. \textbf{179}, 375 (2015).

\bibitem{Schmitt2019} M. Schmitt, D. Sels, S. Kehrein, and A. Polkovnikov, Phys. Rev. B \textbf{99}, 134301 (2019).

\bibitem{Cameo} S. Pilatowsky-Cameo, J. Chavez-Carlos, M. A. Bastarrachea-Magnani, P. Stransky, S. Lerma-Hernandez, L. F. Santos, and J. G. Hirsch, Phys. Rev. E \textbf{101}, 010202(R) (2020).

\bibitem{Nation2019} C. Nation and D. Porras, arXiv:1908.11773.

\bibitem{Wineland1998} D. J. Wineland, C. Monroe, W. M. Itano, D. Leibfried, B. E. King, and D. M. Meekhof, J. Res. Natl. Inst. Stand. Technol. \textbf{103}, 259 (1998).

\bibitem{Schneider2012} C. Schneider, D. Porras, and T. Schaetz, Rep. Prog. Phys. \textbf{75}, 024401 (2012).

\bibitem{Emary2003} C. Emary and T. Brandes, Phys. Rev. E \textbf{67}, 066203 (2003).

\bibitem{Larson2021} J. Larson and T. Mavrogordatos, \emph{The Jaynes-Cummings Model and Its Descendants} (IOP ebooks, 2021).

\bibitem{Larson2008} J. Larson, Phys. Rev. A \textbf{78}, 033833 (2008).

\bibitem{Wang} C. S. Wang \emph{et al}., arXiv:2202.02364.


\bibitem{Deutsch1991} J. Deutsch, Phys. Rev. A \textbf{43}, 2046 (1991).

\bibitem{Garttner2018} M. G\"arttner, P. Hauke, and A. M. Rey, Phys. Rev. Lett. \textbf{120}, 040402 (2018).






\end{thebibliography}
\end{document}